%% file: main.tex
\documentclass[journal]{IEEEtran}

\usepackage{amsmath}
\usepackage{graphicx}
\usepackage{caption}
\usepackage{url}
 \usepackage{multirow}
 \usepackage{multicol}
\graphicspath{{Figures/}}
\usepackage{subcaption}
\usepackage[sort,compress]{cite}
\usepackage{times}
\usepackage{algorithm}
\usepackage{verbatim}
\usepackage{sidecap}
\usepackage{array}
\usepackage{float}
\usepackage{comment}
\usepackage{minibox}
\usepackage{color,soul}
\usepackage{enumerate}

\usepackage{xcolor,soul}
\colorlet{shadecolor}{yellow}
\usepackage{mathtools}
\usepackage{breqn}
\usepackage{multicol}
\usepackage{dblfloatfix}
\input epsf
\usepackage{siunitx}
\DeclareSIUnit\erg{erg}
\DeclareSIUnit\emu{emu}
\DeclareSIUnit\cc{cc}

\input{comments}

\begin{document}
%
\title{A Multi-domain Magneto Tunnel Junction for Racetrack Nanowire Strips}
%
%
%

\author{Prayash Dutta,
        Albert~Lee, 
        Kang~L.~Wang, 
        Alex~K.~Jones, 
        and~Sanjukta~Bhanja
\thanks{P. Dutta and S. Bhanja are with the Department
of Electrical Engineering, University of South Florida, Tampa,
FL 33620, USA e-mail: \{prayash,bhanja\}@usf.edu.}
\thanks{A. Lee and K. L. Wang are with the Department of Electrical and Computer Engineering, University of California at Los Angeles, Los Angeles, CA 90095, USA e-mail: \{alee0618,klwang\}@ucla.edu.}%
\thanks{A.~K.~Jones is with the Department of Electrical and Computer Engineering, University of Pittsburgh, Pittsburgh, PA 15261, USA email: akjones@pitt.edu}
\thanks{This work was partially funded by NSF CNS-1822085, CNS-2133267, CNS-2133340, NSA, and Laboratory of Physical Sciences
.}}

\maketitle
\begin{abstract}
Domain-wall memory (DWM) has SRAM class access performance, low energy, high endurance, high density, and CMOS compatibility.  Recently, shift reliability and processing-using-memory (PuM) proposals developed a need to count the number of parallel or anti-parallel domains in a portion of the DWM nanowire.  In this paper we propose a \textit{multi-domain} magneto-tunnel junction (MTJ) that can detect different resistance levels as a function of a the number of parallel or anti-parallel domains.  
Using detailed micromagnetic simulation with LLG, we demonstrate the multi-domain MTJ, study the benefit of its macro-size on resilience to process variation and present a macro-model for scaling the size of the multi-domain MTJ.  Our results indicate scalability to seven-domains while maintaining a \SI{16.3}{\milli \volt} sense margin.
\end{abstract}


%
\IEEEpeerreviewmaketitle

\vspace{-.1in}
\section{Introduction}
\IEEEPARstart{D}{omain}-wall memory (DWM), or ``Racetrack'' memory~\cite{pieee}, is among the most promising new memory technologies. As a spintronic memory it inherits the SRAM class access performance and low energy of STT-MRAM with a dramatically higher density (as small as $2F^2$, where $F$ is the technology feature size).   Moreover, DWM avoids endurance challenges by providing $\geq10^{16}$ write cycles~\cite{pieee} compared to endurance limited phase-change and resistive memories at $10^{8}-10^{9}$ and $10^{11}-10^{12}$ write cycles, respectively~\cite{NVM-Comparison,pieee}.  


DWMs are formed from ferromagnetic nanowires.  These nanowires extend the magneto-tunnel junction (MTJ) concept of spin-transfer-torque magnetic random access memories (STT-MRAM).  DWM nanowires extend the free-layer to store multiple, \textit{e.g.,} 32-512, magnetically polarized domains that correspond to bits, separated by fabricated notches.   Between adjacent domains storing complimentary bits, a mobile \textit{domain wall} (DW) that pins to these notches balances the exchange and
anisotropic energies~\cite{book94}. Spin-polarized current can shift the magnetic domains with controlled DW motion. 

DWM's improved density comes at the cost of this data shifting, which is necessary to align data with access points.  
There has been significant effort on power reduction and speed improvement~\cite{TapeCache,DWM_Tapestri,pieee} to optimize DWM shifting overhead and others to address shift reliability~\cite{hifi,ollivier2019dsn}.  

\begin{figure}[tbp]
 \centering
\includegraphics[width=\columnwidth]{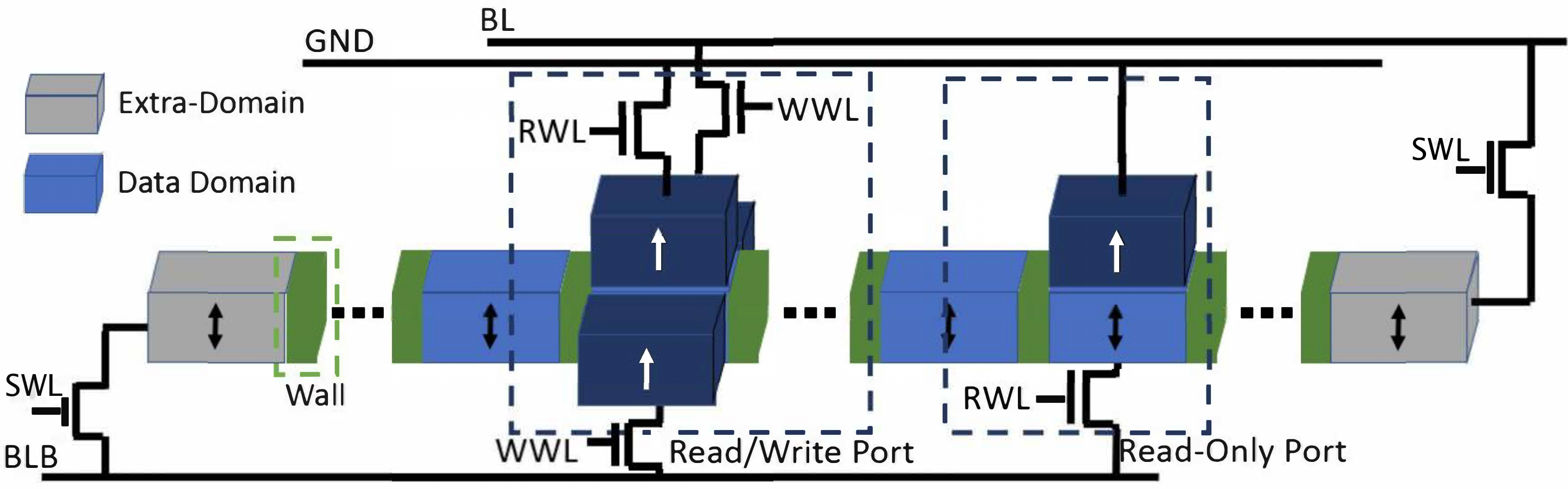}
\caption{Anatomy of a DWM nanowire~\cite{DWM_Tapestri}.}
\vspace{-.2in}
\label{DWMzoom}
\label{fig:nanowire}
\end{figure}

Transverse read~\cite{roxy2020novel} has recently been proposed to count the number of parallel or anti-parallel (\textit{i.e.,} count of `1's) between two access ports of the nanowire.  It applies a much smaller than a shifting current, $R_{T} << R_{S}$ across \texttt{BLB} and \texttt{GND} by opening \texttt{WWL} in Fig.~\ref{fig:nanowire} to detect the tunneling magnetoresistance (TMR) of multiple domains against the fixed layer of an MTJ, \textit{e.g.,} the read-only access port shown in dark blue.  However, proximity of the domain to the tunneling effect can create variation in the resistance for different permutations of data and limit the scalability of how many domains can be sensed.  Furthermore, this structure is likely quite sensitive to process variation.  Thus, we propose a \textit{multi-domain magneto tunnel junction} device that can count the number of 1's in a segment of the nanowire while having the potential for improved scalability and better resiliency to process variation.  

\begin{figure}[bp]
\vspace{-.2in}
 \includegraphics[width=\columnwidth]{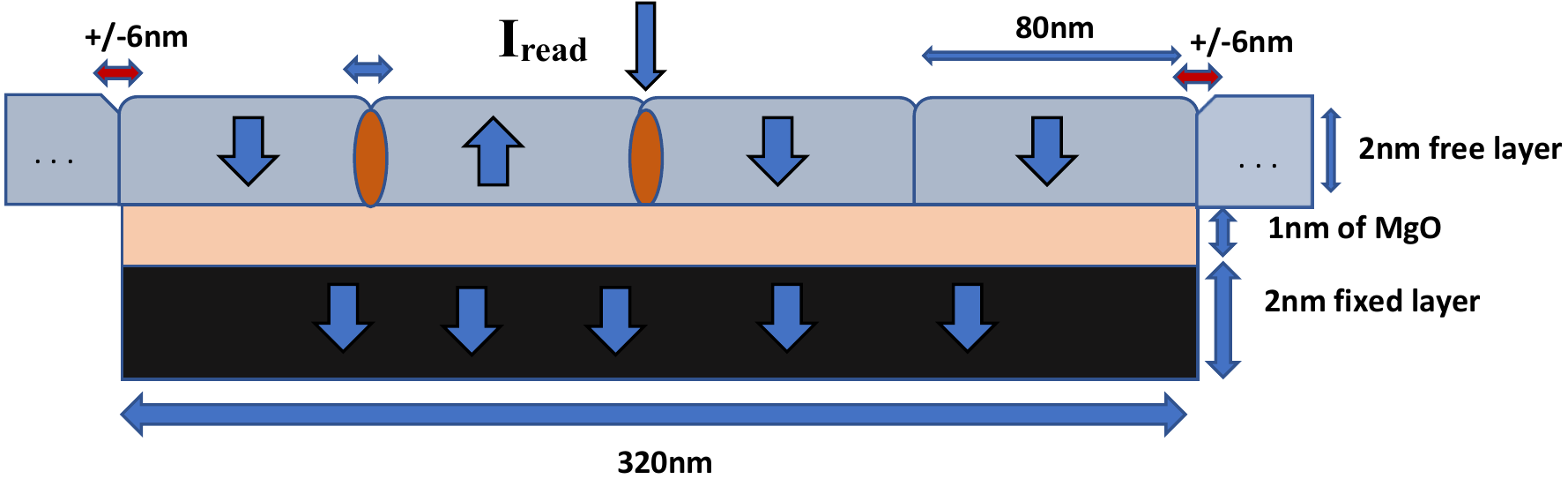}
 \caption{The structure of the experimental setup. The brown ellipses are the domain walls between two oppositely magnetized domains in the free layer.}
 \vspace{-.025in}
 \label{fig:multi-domain-mtj}
 \label{Setup}
 \end{figure} 
\vspace{-.3in}
\section{Multi-domain Magneto Tunnel Junction}
Our proposed multi-domain MTJ is shown in Fig.~\ref{fig:multi-domain-mtj}.  The multi-domain MTJ is similar to the read-only access port from Fig.~\ref{fig:nanowire} except visualized with the fixed layer and MgO below the free layer and covering multiple domains.  
The system behaves like $k$ parallel resistors where $k$ is the number of domains covered by the multi-domain MTJ.  Each domain's region forms a high or low resistance state depending on parallel (-Z) or anti-parallel (+Z) magnetic polarization.  Thus, the multi-domain MTJ resistance is determined by the number of parallel (or anti-parallel) domains, representing how many, but not which domains store `1's.

Prior work has used a spintronic nanowire with a flexible DW to store an analog weight for neural network processing~\cite{RoyAnalogMTJ}.  This requires fine control over the DW motion and high precision sense-amplifiers to detect different analog values.  While there are some smaller effects which do perturb the resistance levels based on the actual pattern of `1's in the device, our multi-domain MTJ uses notches to ensure domain stability while shifting and functions as digital device.  Z-direction current applied evenly across the multi-domain MTJ can induce flow in the +/-X direction to seek lower resistance paths inducing some Anisotropic Magenetoresistance (AMR).  Additionally, DW regions have different resistance properties.  Thus, domains containing ``1010'' will have slightly different resistance than than ``1100'' due to different numbers of DWs.

Because the multi-domain MTJ is much larger than $F$, we expect variation to only impact the device at the extremities, creating a small amount of under- or over-hang of boundary domain notches, while the internal domains are covered in their entirety.  Thus, the impact is less significant than when aligning a single MTJ to a single DWM domain or detecting the resistance with a freely moving DW.


\vspace{-.1in}
\section{Experimental Setup and Results}

\renewcommand{\tabcolsep}{4pt}
\begin{table}
\caption{Multi-domain MTJ Properties and Dimensions}
\label{tab:experiment-parameters}
 \footnotesize
 \begin{tabular}{l|l|l|l}
 \hline \hline
 \multicolumn{2}{l|}{\textbf{Property}} & \multicolumn{2}{l}{\textbf{Values}} \\ \hline
 \multicolumn{2}{l|}{Fixed and Free Layer Material} & \multicolumn{2}{l}{CoFeB}\\ 
   \multicolumn{2}{l|}{Fixed and Free Layer Size} &  \multicolumn{2}{l}{\SI{320}{\nano\meter}$\times$\SI{40}{\nano\meter}$\times$\SI{2}{\nano\meter}}\\ 
    \multicolumn{2}{l|}{Domain Size} & \multicolumn{2}{l}{\SI{80}{\nano\meter}$\times$\SI{40}{\nano\meter}$\times$\SI{2}{\nano\meter}} \\ 
    \multicolumn{2}{l|}{Notch dimensions} & \multicolumn{2}{l}{\SI{12}{\nano\meter}$\times$\SI{10}{\nano\meter}$\times$\SI{2}{\nano\meter}} \\ 
    \multicolumn{2}{l|}{MgO Layer Size} &   \multicolumn{2}{l}{\SI{320}{\nano\meter}$\times$\SI{40}{\nano\meter}$\times$\SI{1}{\nano\meter}}\\\hline
    \multicolumn{2}{c|}{\textbf{CoFeB Parameters}} & $K_u$ & \SI{99999}{ \erg / \cc} 
    \\ \cline{1-2}
    AMR Ratio & $0.014$ & $Ms$ & \SI{1200}{\emu / \cc}   \\ 
    Resistivity & \SI{15}{\micro \ohm \centi \meter} & Exchange Stiffness & \SI{2.2}{\micro \erg / \centi \meter}   \\ \hline
    MgO TMR Ratio & $0.8$ &
   $J_C$ & $3.21\times10^{10}$\SI{}{\ampere / \meter \squared }\\ \hline \hline
 \end{tabular}
\label{table:Parameters}
\vspace{-.175in}
\end{table}
\renewcommand{\tabcolsep}{6pt}

We simulated the multi-layer structure from Fig.~\ref{fig:multi-domain-mtj} using the LLG micromagentic simulator~\cite{LLG} using CoFeB and MgO with the parameters shown in Table~\ref{tab:experiment-parameters}, where $K_u$ is the uniaxial anisotropy 
and $Ms$ is the saturation magnetization.  Domains were of \SI{80}{\nano \meter} long, \SI{40}{\nano \meter} wide, and \SI{2}{\nano \meter} thick.  4 domains were included in the MTJ with a \SI{1}{\nano \meter} thick MgO layer and \SI{2}{\nano \meter} thick CoFeB fixed layer. 
Notches of size \SI{12}{\nano \meter} $\times$ \SI{10}{\nano \meter} $\times$ \SI{2}{\nano \meter}  create pinning sites for DWs.  

The fixed layer is magnetized in the -Z direction.  Each free layer domains can be magnetized in either +/-Z. The MgO layer is a non-magnetic tunneling barrier.  We assign +Z (anti-parallel, high resistance) as `1' and -Z (parallel, low resistance) as `0'.
Our read current ($I_{read}$) has a read current density $J_{c}= 3.21 \times 10^{10} A/m^{2}$ and remains an order of magnitude lower than the switching current density to minimize the potential of read disturbance~\cite{kang2021critical}.  For more than four domains, $I_{read}$ can increase to keep $J_C$  invariant.  We attempted to match experimental parameters with prior transverse read work~\cite{roxy2020novel}.

\begin{figure}[tbp]
\vspace{-.1in}
\center 
\includegraphics[width=\columnwidth]{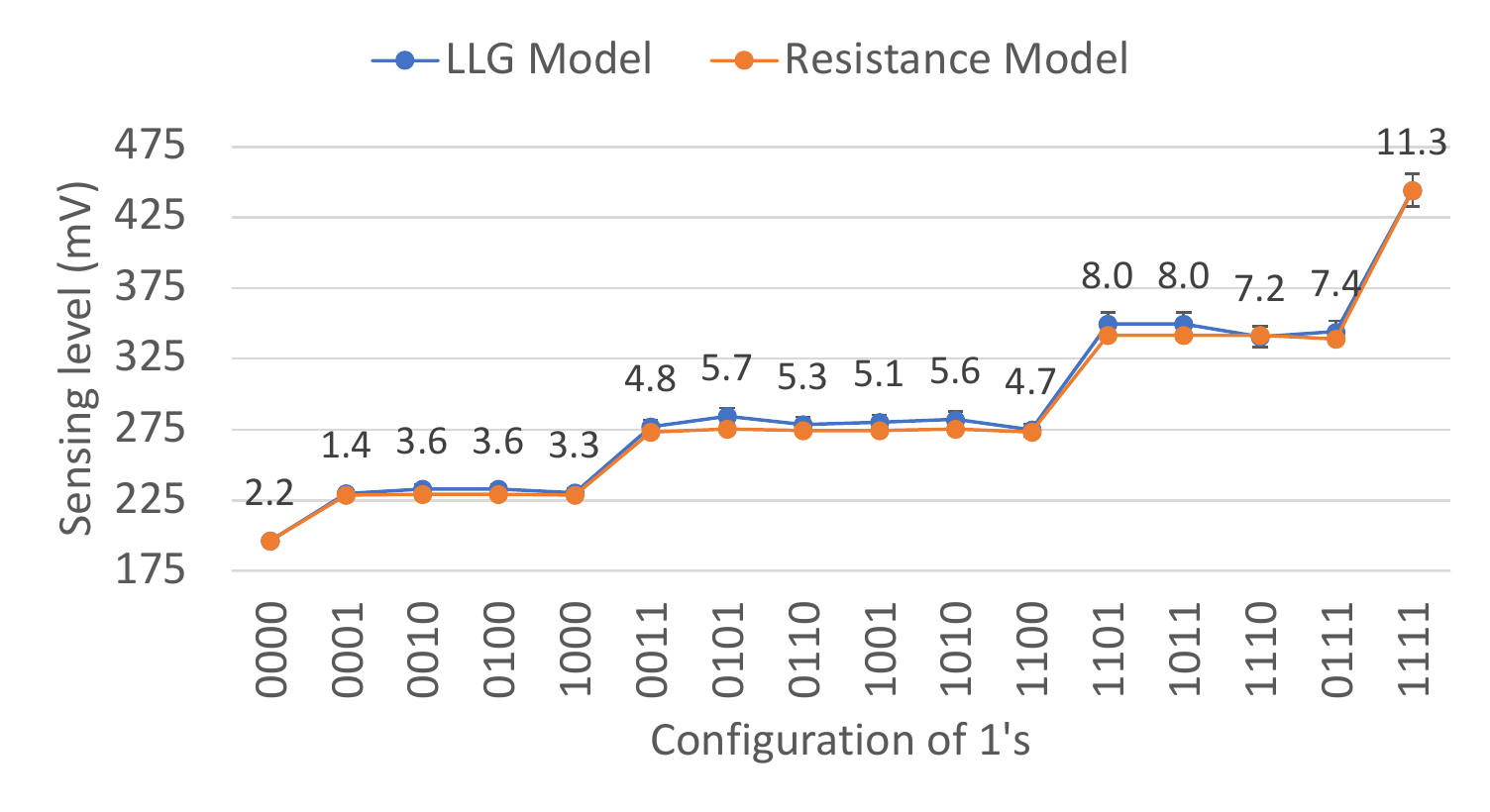}
\vspace{-.2in}
\caption{Sense margin for all permutations of four domains in the free layer.  Error bars and labels report 6$\sigma$ error deviation.}
\vspace{-.25in}
\label{resistances}
\end{figure}

The resistance is measured from the top of the free layer to the bottom of the fixed layer in the simulator which originates TMR effects from the MTJ and AMR effects from DWs. 
The impact of domain contents on voltage is shown graphically in blue based on micromagnetic simulation in Fig.~\ref{resistances}.  We see good similarity of the voltages for all combinations with the same `1's count and the margin between clusters grows as the number of `1's increases.  Variability in the cluster is primarily due to the number of DWs.
The minimum margin is is \SI{33.5}{\milli \volt} between ``0000'' and ``0001'' and the margin grows between groups with more `1's.

To consider process variation we simulated the same model with a \SI{2}{\nano \meter} and \SI{6}{\nano \meter} misalignment between the notch in the extended free layer and the MgO and CoFeB fixed layer (red arrows in Fig.~\ref{fig:multi-domain-mtj}), where we treated \SI{5.5}{\nano \meter} as the $6\sigma$ in a standard distribution of variation~\cite{komalan2017cross}.  In this structure, the MgO and CoFeB fixed layer alignment would typically be fabricated with trenches and etching followed by chemical vapor deposition of doped CoFeB and MgO.  The alignment concern would come from the location of the notch with the trench wall.  The sense margin deviation for \SI{6}{\nano \meter} is shown with error bars to the blue series in Fig.~\ref{resistances}.  As the error bars are small we also report this deviation in \SI{}{\milli \volt} as labels.  
From our read current density, the lowest voltage level that has to be recognized for this resistance gap is still a healthy \SI{28.4}{\milli \volt} which is practical for conventional sense amplifiers \cite{salehi2017survey}. 
\begin{table*}[tbp]
\begin{equation}
J_CDA_d \left\{  \left( \frac{D - 2}{R^-_{80}} + \frac{1}{R^-_{74}} + \frac{1}{R^+_{68}} + \frac{1}{R^{DW}_{0\rightarrow1}}  +  \frac{1}{R^{hDW}_{+Z}} \right)^{-1} - \left(\frac{D - 2}{R^-_{80}}+\frac{2}{R^-_{74}}+\frac{2}{R^{hDW}_{-Z}}\right)^{-1} \right\}
\label{eq:scaling-margin}
\end{equation}
\vspace{-.2in}
\end{table*}


\vspace{-.1in}
\section{Analytical Model and Scaling}
\begin{figure}[tbp]
\vspace{-.1in}
\center 
\includegraphics[width = \columnwidth]{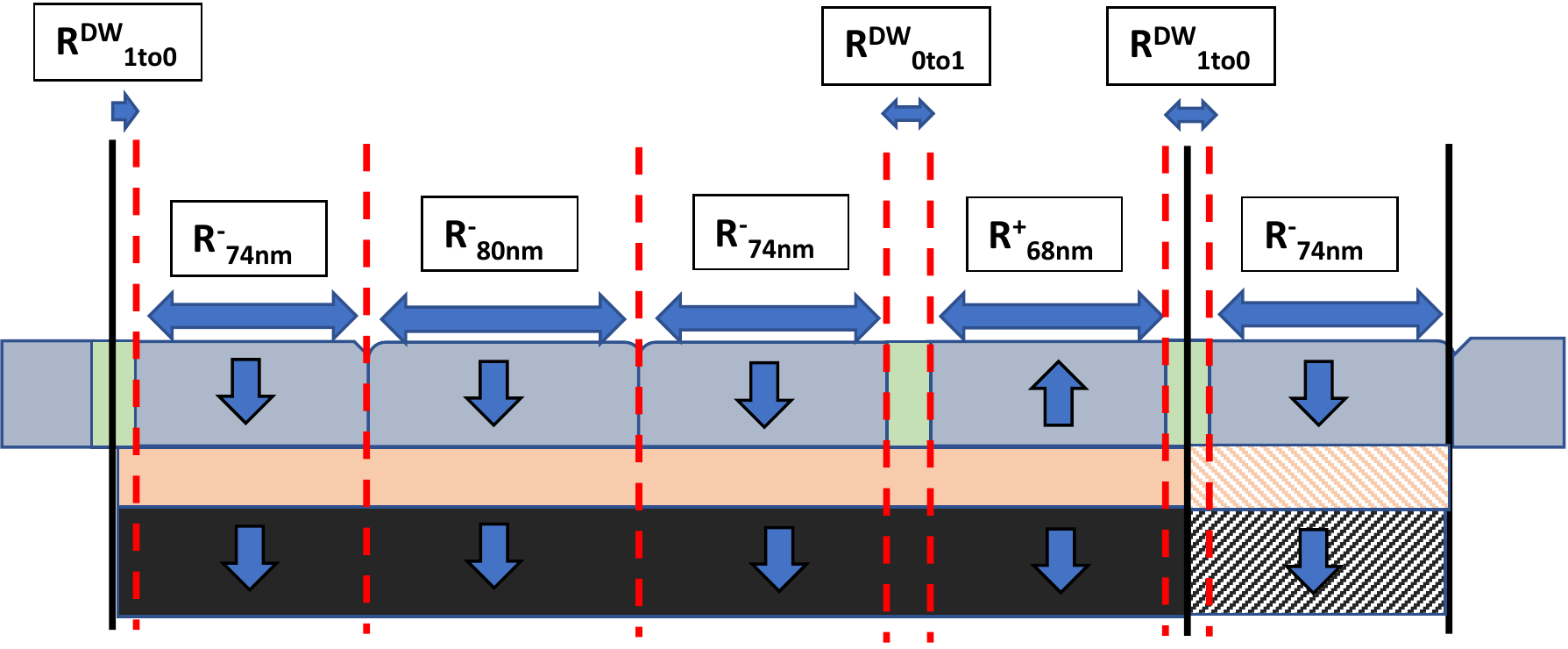}
\caption{Resistance modeling of domains in the free layer}
\label{analytical}
\vspace{-0.1in}
\end{figure}

Detailed magnetic modeling with LLG becomes impractical as number of domains scales.  Like prior approaches for  analog DW movement in a nanowire~\cite{WangAnalogMTJSpice} and spintronic logic proposals~\cite{FriedmanMTJLogicSpice}, we have constructed a characterized analytical model of the multi-domain MTJ which we describe in Fig.~\ref{analytical}. This entire model is a parallel arrangement of some mini-resistive structures.  The 4-domain example shown in the figure shows a ``0001'' but both bordering values are different.  Thus, the leftmost domain has a low resistance configuration, denoted R$^-$ for \SI{74}{\nano \meter} and half a DW.  The next domain has R$^-$ for \SI{80}{\nano \meter}, followed by R$^-$ for \SI{74}{\nano \meter}, a full DW, and R$^+$ for \SI{68}{\nano \meter}, with another half DW.  From characterization these can be converted to particular resistance values shown in Table~\ref{tab:mini-structures}.  We can scale to a five-domain MTJ by adding the shaded fixed and barrier layers in Fig.~\ref{analytical} modeling ``00010.''  The expression replaces the rightmost half DW with a full DW and another R$^-$ for \SI{74}{\nano \meter}.  These divisions are segmented by red dotted lines  with black lines indicating the borders of the 4-domain and 5-domain examples.  ``00010'' can be calculated through the parallel equivalent resistance to be \SI{431.5}{\ohm}.

\renewcommand{\arraystretch}{1.4}

\renewcommand{\tabcolsep}{2pt}
\begin{table}[tbp]
\centering
\footnotesize
\centering
\caption{Simulation mini structures and resistance}
\label{tab:mini-structures}
\begin{tabular}{|l|l|l|}
\hline
Direction: Size &Notation&Resistance (\SI{}{\ohm})\\ \hline
\textbf{-Z}: \SI{80}{\nano \meter}, \SI{74}{\nano \meter}, \SI{68}{\nano \meter} &$R^{-}_{80}$, $R^{-}_{74}$, $R^{-}_{68}$& 1911, 2048, 2228 \\ \hline
\textbf{+Z}: \SI{80}{\nano \meter}, \SI{74}{\nano \meter}, \SI{68}{\nano \meter} &$R^{+}_{80}$, $R^{+}_{74}$, $R^{+}_{68}$& 4324, 4730, 5143 \\ \hline
\textbf{DW} -Z to +Z, +Z to -Z: \SI{12}{\nano \meter} & $R^{DW}_{0\rightarrow1},R^{DW}_{1\rightarrow0}$&20053, 20063\\ \hline
\textbf{Half DW} -Z to Z$_\O$, +Z to Z$_\O$: \SI{6}{\nano \meter} &$R^{hDW}_{-Z}$, $R^{hDW}_{+Z}$& 
35061, 46196\\ \hline
\end{tabular}
\label{structure}
\vspace{-.15in}
\end{table}
\renewcommand{\tabcolsep}{6pt}

\begin{figure}[tbp]
\vspace{-.25in}
\center 
\includegraphics[width=\columnwidth,scale=0.75]{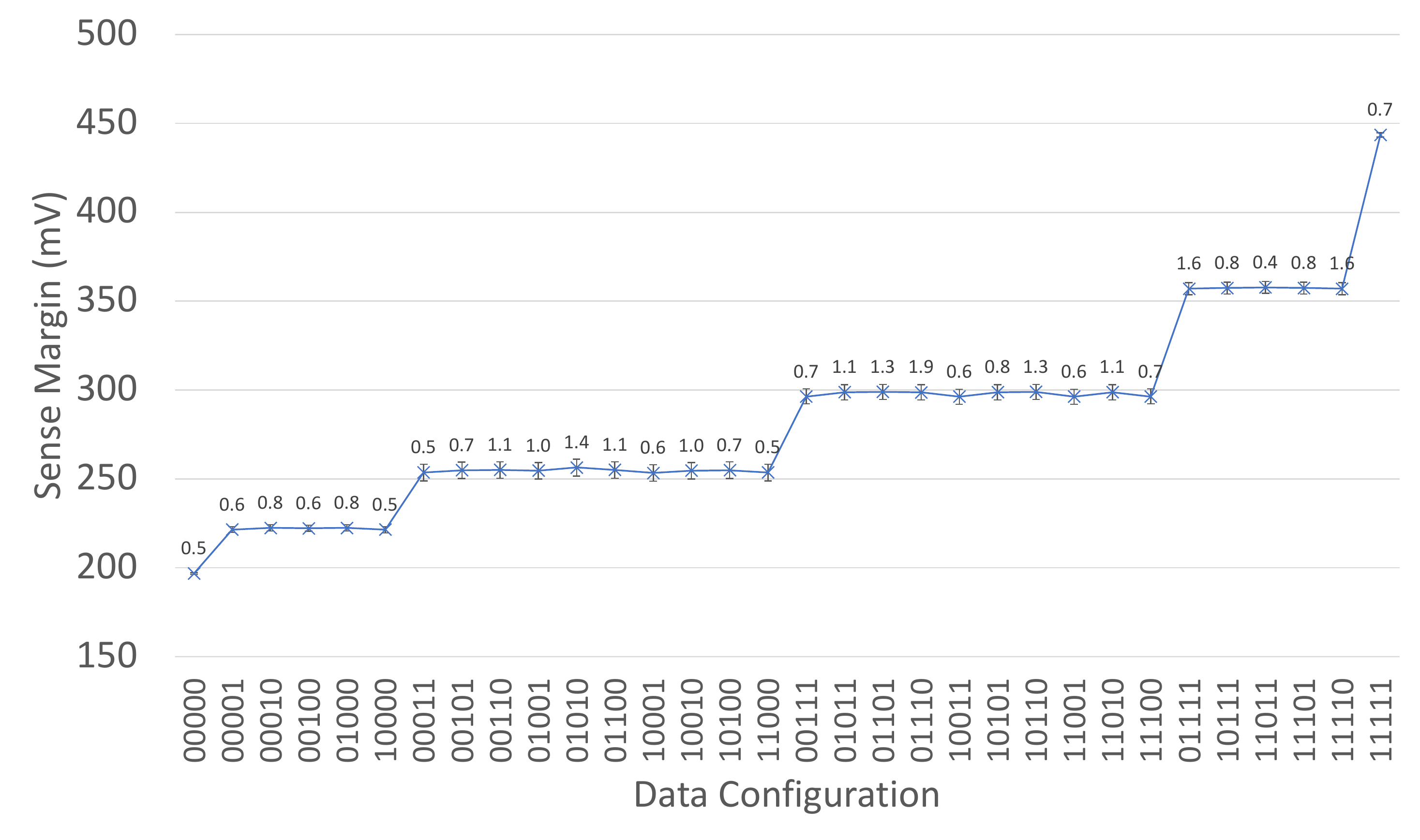}
\vspace{-.2in}
\caption{Modeled 5-domain MTJ, labels report variation due to neighboring domains}
\label{fivebitgraph}
\vspace{-.05in}
\end{figure} 

We demonstrate accuracy of matching this model to the micromagnetic simulation data for a four domain MTJ in the orange series of Fig.~\ref{resistances}, which has an average and maximum error of 1.4\% and 3.2\%, respectively.  We also show data from the five-bit domain model with resistances presented in Table~\ref{fivebit} and sensing margins presented in Fig.~\ref{fivebitgraph}.  We checked data for different numbers of `1's through micromagnetic simulation and determined an average error of $<1\%$.  This data demonstrates six different distinguishable levels when applying the same $J_C$ projects a minimum sense margin is \SI{23.7}{\milli \volt}, which remains detectable by standard sense amplifiers~\cite{salehi2017survey}.  We can represent the minimum sense margin as described in Eq.~\ref{eq:scaling-margin} where $D$ is the number of domains in the multi-domain MTJ, $A_d$ is the area per domain, and values of $R$ are found in Table~\ref{tab:mini-structures}.  For six and seven domains, the minimum margin scales to a still detectable \SI{19.3}{\milli \volt} and \SI{16.3}{\milli \volt}, respectively.

\begin{table}[tbp]
\vspace{-.1in}
\centering
\caption{Projected Five-domain Resistances}
\begin{tabular}{|l|l||l|l|}
\hline

Combination&Resistance&Combination&Resistance\\   \hline
\multicolumn{2}{|l||}{\hspace{0.3in}All 0's $\downarrow$} &\multicolumn{2}{l|}{\hspace{0.3in}Three 1's $\downarrow$} \\   \hline
00000&382.10&00111, 11100&576.22\\   \hline
 \multicolumn{2}{|l||}{\hspace{0.3in}One 1's $\downarrow$} &01011, 11010&580.11\\   \hline
00001, 10000 &431.07&01101, 10110&581.07\\   \hline
00010, 00100, 01000 &431.50&01110, 10011, 11001&577.94\\   \hline
\multicolumn{2}{|l||}{\hspace{0.3in}Two 1's $\downarrow$} &10101&583.27\\   \hline
00011, 11000 &493.19& \multicolumn{2}{l|}{\hspace{0.3in}Four 1's $\downarrow$} \\   \hline
00101, 10100 &496.03&01111, 11110&692.87 \\   \hline
00110, 01100, 10001 &494.45 &10111, 11011, 11110 &697.39  \\   \hline
01001, 10010 &495.01 &  \multicolumn{2}{l|}{\hspace{0.3in}Four 1's $\downarrow$} \\ \hline
01010&496.6 &11111&864.88\\   \hline
\end{tabular}
\label{fivebit}
\vspace{-.1in}
\end{table}


\vspace{-.1in}
\section{Conclusion}
In this paper, we have proposed a brand new technique to count the number of `1's in a DWM nanowire segment.  The multi-domain MTJ is sufficiently resistant to effects from process variation.  In addition we have presented an analytical model which can be used to explore different configurations of fewer or more domains.  
In our future work we hope to explore extending the analytical model with impacts from process variation, as well as ways to increase the sense margin.
\vspace{-.1in}


%





\ifCLASSOPTIONcaptionsoff
  \newpage
\fi



%

\bibliographystyle{ieeetr}
\bibliography{bib/dwm}
%








\end{document}

%% file: comments.tex
\usepackage{ifthen}
\usepackage{amssymb}
\usepackage{xcolor}
\usepackage{comment}

\newboolean{showcomments}
\setboolean{showcomments}{true}

\makeatletter
\newcommand{\mynote}[3]{%
  \ifthenelse{\boolean{showcomments}}{%
   \fbox{\bfseries\sffamily\scriptsize#1}%
   {\small$\blacktriangleright$\textsf{\emph{\color{#3}{#2}}}$\blacktriangleleft$}}%
  {%
   \@bsphack
   \@esphack
  }%
}
\makeatother

\definecolor{asparagus}{rgb}{0.53, 0.66, 0.42}